\documentclass[%
 reprint,showkeys,amsmath,amssymb,
 aps,pra,onecolumn,floatfix]{revtex4-2}

\usepackage{graphicx}
\usepackage{dcolumn}
\usepackage{bm}
\usepackage{natbib}
\usepackage{color}
\usepackage{gensymb}

\newcommand{\Xe}{{$^{129}\mathrm{Xe}$}}
\newcommand{\Rb}{{$^{87}\mathrm{Rb}$}}
\newcommand{\He}{{$^{3}\mathrm{He}$}}
\newcommand{\PC}{{PANCo}}
\newcommand{\SCC}{{SCC}}
\newcommand{\pn}{{$\mathbf{P^n}$}}
\newcommand{\pe}{{$\mathbf{P^e}$}}

\begin{document}

\title{Dual axis atomic magnetometer and gyroscope enabled by nuclear spin perturbation}

\author{Morgan Hedges}
\author{Ankit Papneja}
\author{Karun Paul}
\author{Ben C Buchler}
\email{ben.buchler@anu.edu.au}

\affiliation{%
Research School of Physics\\
Australian National University\\
Acton 2601, Australia
}
%\date{\today}

\begin{abstract}
Alkali-noble-gas comagnetometers have become an essential tool for tests of fundamental physics and offer a compact platform for precision gyroscopy. They are, however, limited by technical noise at low frequencies, commonly due to their limited suppression of magnetic noise. Here we investigate a new method for co-magnetometry between a single noble gas and alkali species. While similar to well-known devices using self-compensation, our scheme introduces magnetic pulses that controllably perturb the noble gas and pulsed optical pumping to polarise the alkali atoms. These applied pulses allow our scheme to measure, rather than just suppress, the effect of magnetic noise thereby offering reduced cross-talk.  We show numerically that our scheme retrieves four signals (rotations and magnetic fields on two transverse axes) with similar sensitivity to a single axis device. We also present a proof-of-principle experiment based on a \Rb{}-\Xe{} cell. Our data shows a low magnetic-rotation cross-talk of $0.2 \pm 0.1\mu$Hz$/$pT, which is already on par with the most sensitive devices relying on self-compensation.
\end{abstract}


\maketitle

\section{Introduction}

Atomic sensors based on spin polarised ensembles have long been used for precision measurement. This includes measurements of magnetic fields, rotation, tests of fundamental physics \cite{Budker,Terrano_2022,PhysRevA.108.010101}, medical imaging \cite{BROOKES2022621} and device characterisation \cite{Hu2020}.
In all these applications, an alkali species is optically pumped to polarise the spin. The precession of this ensemble can then be measured with an optical probe.

The large gyromagnetic ratio of alkali species means that alkali ensembles are particularly excellent sensors of magnetic fields. Indeed, modern atomic magnetometers based on spin polarised alkali atoms reach state of the art sensitivity, comparable to to Superconducting Quantum Interference Devices (SQUIDs) \cite{Degen2017,Mitchell2020}. An advantage over SQUIDs is that atomic magnetometers do not require cryogenic cooling, enabling cheaper and wider deployment in applications like magnetoencephalography \cite{BROOKES2022621}.  The highest sensitivity atomic magnetometers rely on the Spin-Exchange Relaxation-Free (SERF) regime \cite{Allred2002}, which requires operating in a low-field environment. It has also been shown that full 3-axis readout is possible with SERF magnetometers via the application of magnetic field modulations \cite{10.1063/1.1814434,Yan:22}.

% Atomic gyroscopes

Atomic gyroscopes on the other hand, promise sensitivity comparable to the best optical systems, but with smaller volumes and lower power consumption. The gyromagnetic ratio of alkali atoms is problematic here though: they respond much more strongly to magnetic fields than any other effect.

The best known method to improve non-magnetic sensing in these systems is to introduce a noble-gas with a long-lived nuclear spin into the same cell to make a `comagnetometer' \cite{Kornack2002,Kornack2005}. Optical pumping of the alkali electron spin polarisation, \pe{},  pumps the noble gas nuclear spin polarisation, \pn{}, via spin-exchanging collisions. The polarised noble gas has three important properties.
First, the gyromagnetic ratio is up to 1000 times smaller than that of alkali atoms, so it has a much smaller response to magnetic noise.
Second, due to the much ($>10^4 \times$) higher number density of the noble gas, the effective magnetisation of \pn{} is typically much stronger than that of \pe{}.
Third, the coherence- and life-times of \pn{} (seconds to hours) are much longer than those of \pe{} ($\sim$ms).
This combination of properties mean that \pn{} will provide a strong, stable  polarisation that is less sensitive to magnetic fields and thus better suited to sensing non-magnetic effects.
What it lacks, however, is a suitable optical transition that allows direct measurement of \pn{}. In a comagnetometer, we use the fact that \pn{} provides strong magnetic driving of \pe{}. Provided external magnetic noise can be sufficiently suppressed, an optical measurement of \pe{} may provide a precise readout of \pn{}.

To distinguish the target signal from external magnetic noise, a common method is the Nuclear Magnetic Resonance (NMR) gyroscope which uses two noble gas species with different nuclear gyromagnetic ratios \cite{WALKER2016373}. The two species make nominally independent measurements of the effective magnetic fields they see, and a discrepancy in the measurements reveals the rotation rate. The role of the alkali atoms in this scheme is to pump the noble gas species and measure their dynamics, via the magnetic interaction between \pn{} and \pe{}. Recent work has also shown that modulation of the pump and applied bias field in NMR systems may be useful for reducing technical noise \cite{PhysRevA.100.061403,s23104649}.
The alkali in an NMR gyroscope does not work in the SERF regime, however, meaning that spin relaxation noise limits how well an alkali species can measure the precession of the noble gas. To partly mitigate spin relaxation, pump pulses with alternating polarisation \cite{Limes2018} and magnetic pulses \cite{Korver2015}  have shown promise.

The scheme that we have developed is most closely related to the `self-compensated comagnetometer' (\SCC{}) \cite{Kornack2002,Kornack2005,Smiciklas2011} where co-magnetometry is between a single noble gas species and the alkali itself. In the SCC, a magnetic bias field is applied along the pump axis. The bias is tuned to precisely counteract the combined magnetisation of the alkali and noble gas spins in the cell. Under these conditions it will also be the case that the Larmor frequencies of the alkali and noble gas will be near-equal so the ensembles are strongly coupled. Under these conditions it can be shown that low frequency magnetic noise drives \pn{} in a way that leads to cancellation of the external field, so that magnetic noise on \pe{} is suppressed. 

Non-magnetic influences such as rotation, on the other hand, will not be suppressed and can even be enhanced by the strong coupling between species. A significant advantage of the \SCC{} is that it can use high density alkali in the SERF regime, improving the sensitivity of the readout. This technique has been instrumental in setting new limits in various anomalous physics searches, including searches for a 5th force \cite{PhysRevLett.125.201802}, violations of Lorentz invariance \cite{Smiciklas2011} and dark matter \cite{PhysRevD.97.055006,Bloch2020,Padniuk2022}.

The gyroscopic performance of the \SCC{} has shown excellent levels of sensitivity. For example Kornack et al. \cite{Kornack2005} found a noise floor $\mathrm{5\!\times\!10^{-7}\,rad\,s^{-1}\,Hz^{-1/2}}$, which was recently surpassed by Wei et al. \cite{PhysRevLett.130.063201} with $\mathrm{3\!\times\!10^{-8}\,rad\,s^{-1}\,Hz^{-1/2}}$.
While impressive, these results are still over an order of magnitude above the atomic shot noise limit set by the alkali, and particularly limited at low frequencies where magnetic noise is typically largest. While some progress has been made reducing magnetic and other technical noise sources \cite{8712526,9663281,Wang2022APS,Wang2024phd,Jiang2022APL,PhysRevLett.133.023202}, a remaining challenge is that increased suppression of magnetic noise is not only hard to achieve, it is also difficult to maintain this suppression due to drifts in the operating point.

In this work, we introduce a Perturbed Alkali-Noble Comagnetometer (PANCo) scheme. It may be considered an evolution of a \SCC{}, although self-compensation is no longer required.
\begin{figure}[b!]
    \centering
    \includegraphics[width=0.7\linewidth]{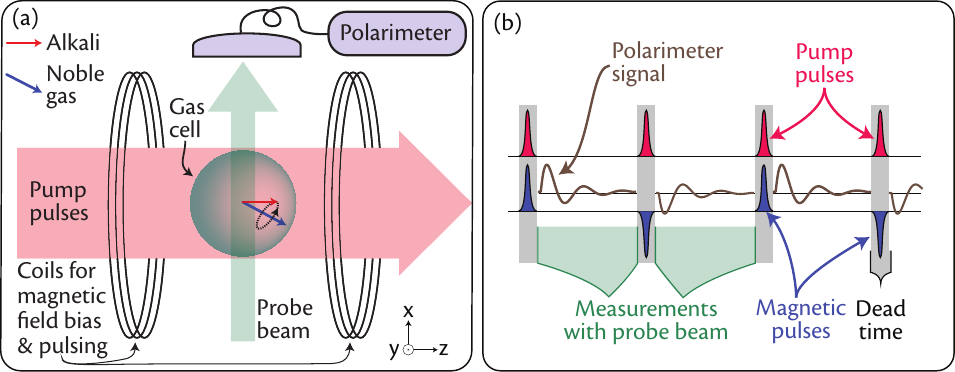}
    \caption{a) The essential components of the new scheme. A gas cell containing a mixture of noble gas and alkali metal is periodically pumped by strong laser pulses along the z-axis. This pumping creates a strong average alkali polarisation along z, which in turn also polarises the noble gas via spin exchanging collisions. A probe beam along the x-axis reads out the x component of the alkali spin. A perturbation of the system by a magnetic field or rotation means that the noble gas polarisation is tilted away from the z-axis. In this case, after excitation with the pump pulse, the alkali polarisation will precess around the total magnetic field in the cell, that will be a combination of the noble gas magnetisation, any external fields and the bias field applied along the z-axis. This leads to a signal on the polarimeter to be analysed. b) The pulse sequence used in our scheme. The $\pi/2$ magnetic pulses rotate the noble gas polarisation around the z-axis with alternating phase, causing any deviation of the noble gas away from z will cause a modulated field seen by the alkali. The optical pump pulses polarise the alkali atoms along the z axis. The precession of the alkali atoms is measured in the the intervals between pulses.}
    \label{fig:scheme}
\end{figure}
The essential changes are the use of synchronous optical pump and magnetic pulses along the $z$-axis, as illustrated in Fig.~\ref{fig:scheme}. The magnetic pulses perturb \pn{}, yet largely preserve its long-term average evolution. However the magnetic pulses do not affect \pe{} as it is reset at the start of each measurement cycle by the optical pump.  We will show that the controlled perturbation of \pn{} enables our scheme to separate the influence of magnetic and non-magnetic driving.  Meanwhile the pulsed optical pump and readout of each period of free precession of \pe{} reduces the impact of laser intensity noise and means that we also gain sensitivity along the two axis perpendicular to the pump \cite{Wang2022APS,Wang2024phd}.

%The ability to separate magnetic field signals means that, in combination with pulsed pumping and readout, 
The net result is that our scheme provides four output channels consisting of two rotational and two magnetic signals along axes perpendicular to the pump.  We note that pulsed pumping \cite{Wang2022APS,Wang2024phd} and magnetic modulation \cite{Li_2006,Jiang2022APL,PhysRevApplied.17.024004} have been used previously to achieve dual-axis readout of rotations. Our work is, to the best of our knowledge, the first that combines dual-axis magnetic and rotational readout in a comagnetometer device. By transforming magnetic fields from a noise source into a distinguishable signal, the scheme can further reduce contamination of the rotation signal by magnetic noise. Our scheme also works over a wide range of bias field values, not just at the self-compensation point. The pulsed readout of the system also allows monitoring and suppression of other noise sources \cite{Wang2022APS,Wang2024phd}, such as pointing error.

\section{Methods}

% Geometry/setup components
The requisite components to implement PANCo are in common with an \SCC{} and shown in Fig.~\ref{fig:scheme}(a). A gas cell containing a mixture of alkali metal and noble gas is heated to be in the SERF regime. The $z$-axis of the system is defined by the pump light that gives the alkali a high average polarisation in the $z$ direction. Over time, this will cause a significant polarisation of the noble gas nuclei along $z$,  via spin-preserving collisions. Once fully pumped, the polarised noble gas gives rise to an effective magnetic field of $\sim100\,$nT within the cell. To observe the evolution of the polarisation of the alkali atoms we have a probe beam that is aligned along the $x$-axis. Through the off-resonant Faraday interaction, this beam reads out the $x$-component of \pe{}. 

% Bias field here and vs SCC
In an \SCC{}, the z-axis coils are used to apply a bias field opposite to the species magnetisation, to reach the `compensation point' or `self-compensation' regime. In our scheme we also apply a bias, but here we consider it a parameter. We will see later that operation near the compensation point can improve sensitivity, but it's not a requirement. 

% Description of pulsed pumping
The first main experimental change from a standard continuous wave (CW) \SCC{} is the use of pulsed optical pumping. These pulses aim to rapidly (${\sim}10\,\mu$s) saturate the alkali, with a period of free precession afterwards while the signal is collected. The time between pulses, $\tau$, is on the order of the alkali lifetime ($\sim$ms), such that the average alkali polarisation is similar to 50\% as in the CW case. While using pulses adds some complexity, advantages include a reduction in laser pointing and intensity noise as well as yielding a second measurement axis \cite{Wang2022APS,Wang2024phd}.

% Description of magnetic pulses
The other main difference to a CW \SCC{} is the addition of magnetic pulses along the $z$-axis. These pulses rotate the noble gas about the $z$-axis and are applied simultaneously with the optical pump pulses. 
This ensures that the alkali atoms are held parallel to $z$ so that they are not magnetically driven. It is these pulses that allows the separation of the noble gas and alkali evolution, and thus separation of magnetic fields from rotations. 
In the implementation shown here, the pulses are tuned to give a $\pi/2$ rotation for the noble gas and alternate in sign, although we note that principles of the scheme apply to a range of pulse parameters, not just those used here. 
In this case, we consider a single measurement cycle to be two adjacent pulse-decay intervals, as shown in Fig.\ref{fig:scheme}(b).

% Putting it all together
Before diving into a model of our scheme, it is helpful to have a picture of how it works. 
Without application of magnetic pulses, consider a constant pure rotation in the $x$-axis. Both the alkali and noble gas species rotate together. The optical pump will reset \pe{} to the $z$-axis and it will then precess around \pn{}, which will remain displaced from the $z$-axis. The direct effect of the rotation on \pe{} may be neglected as the large gyromagnetic ratio of the alkali ensures that its magnetic response to \pn{} is dominant.  Now consider the case of a pure magnetic field along the $x$-axis. This will also result in a movement of \pn{}. When the alkali is pumped back to the $z$-axis, however, \pe{} will now be driven by the combination of \pn{} and the external magnetic field. 
Despite the different origins, rotation and magnetic driving both lead to precession of \pe{} around some effective magnetic field and there is no way to discern the origin of that effective field.

Now consider what happens if we add magnetic pulses. They will force \pn{} to alternate between two distinct polarisations due to the alternating sign of the pulses. For a pure rotation, the precession will, as before, be determined by \pn{}. In the case of a pure external magnetic field, however, the alternating positions of \pn{} combined with the external field will give evolution of \pe{} that is  distinct to that due to rotation. The evolution of \pe{} therefore depends on whether the external signal is a rotation or magnetic field. On top of this, the pulsed readout then also provides separation of signals in the $x$ and $y$ directions, providing four output channels in total for the scheme.

To see what the signals look like and how distinct they are, we will use a model based on coupled equations of motion for the two spin species.

\subsection{Model}

To model our method we can use the coupled Bloch equations for the spin polarisation of the alkali atoms, $\mathbf{P^e}$, and the noble gas atoms, $\mathbf{P^n}$. The superscripts reflect the fact that the alkali atom spin is dominated by the electron spin while the noble gas polarisation is due to its nuclear spin. The equations of motion can be written as \cite{Kornack2005}

\begin{eqnarray}
\frac{d \mathbf{P^e}}{d t} &= \frac{1}{q}\bigg[\gamma_e (\mathbf{B}+\lambda M^n\mathbf{ P^n)\times P^e}
        +R_{se}^{ne}\mathbf{P^n} +{\mathbf{(P^e)}}R_p- ({R_{c}^e}+R_{se}^{en})\mathbf{P^e} \bigg]+\boldsymbol{\Omega}\times \mathbf{ P^e},\nonumber \\ 
\frac{d \mathbf{P^n}}{d t}& =\gamma_n \mathbf{(B}+\lambda M^e\mathbf{ P^e)\times P^n}+ \boldsymbol{\Omega}\times \mathbf{P^n}+ R_{se}^{en}\mathbf{P^e}- (R_{se}^{ne}+R_{c}^n)\mathbf{P^n},
\label{eq:BHE_1}
\end{eqnarray}

where $\mathbf{B}$ is an external magnetic field, $\boldsymbol{\Omega}$ the angular velocity of rotation, $\gamma_e$ is the electron gyromagnetic ratio of the alkali metal,  $\gamma_n$ is the nuclear gyromagnetic ratio of noble gas.  Adding to the effect of the magnetic field there is also the magnetisation of the atomic species. The motion of $\mathbf{P^e}$ and $\mathbf{P^n}$ are driven by $\lambda M^n\mathbf{ P^n}$ and $\lambda M^e\mathbf{P^e}$ respectively  where $\lambda$ is the coupling enhancement between the polarisations, and  $M^{(n,e)}$ are the maximum magnetisations of the atomic species.  Terms that lead to decay of the polarisations are governed by $R_{se}^{(en, ne)}$ and $R_{c}^{(n,e)}$.  These describe the rates of polarisation transfer from alkali to noble gas and noble gas to alkali as well as the spin-destroying collisions experienced by the noble gas and alkali species.  Growth in the polarisation is given by the pumping rate, $R_p$, which results from the pump beam propagating along the $z$-axis. In the SERF regime there is a slowing down factor, $q$ that results from the nuclear angular momentum \cite{Allred2002}.  The slowing-down factor is a function of the polarisation of the alkali atoms.  As our system is pulsed, $q$ will be varying throughout the readout phase as the alkali polarisation decays.

\subsection{Magnetic and rotational response.}
\begin{figure*}[hbtp]
    \centering
    \includegraphics[width=12cm]{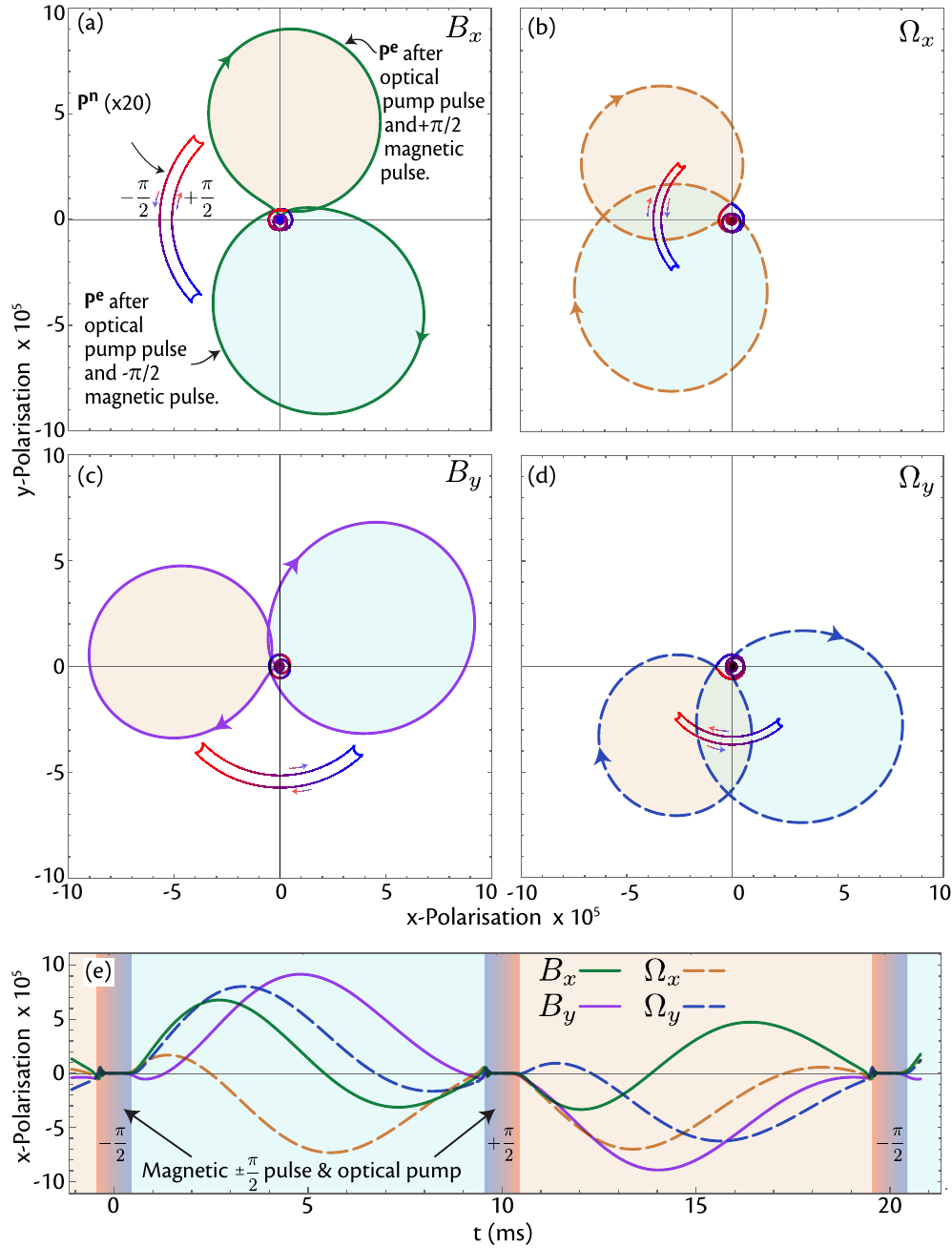}
    \caption{Evolution of the noble gas (\pn{}) and alkali (\pe{}) polarisations in a simulation of \Xe{} and \Rb{}.  Parts (a-d) show the precession of \pe{} and \pn{} projected onto the $x-y$ plane. Part (e) shows the $x$ component of \pe{} that would be measured using the probe beam. The evolution is plotted for a full measurement cycle, consisting of two free precession periods of \pe{} that follow negative and positive magnetic pulses and optical pumping.  The shading in all parts of the figure indicate the phase of the measurement cycle, pale blue after a negative magnetic pulse and pale orange after a positive magnetic pulse.
    Parts (a-d) show that the $x-y$ precession paths for \pe{} are unique for each of the rotations and magnetic fields. During the precession of the the alkali atoms, \pn{} is nearly stationary, reacting only slightly to \pe{} and the external fields. Part (e) shows that the $x$ component of \pe{} is also unique for each of the rotational and magnetic signals. 
    The size of the applied rotations and magnetic fields were chosen to give similar signals in \pe{}, with $B_{(x,y)}$=1.43pT and $\Omega_{(x,y)}$=269$\mu$Hz. Other simulation parameters: \Xe{} Magnetisation: 58nT, \Rb{} Magnetisation: 54nT, bias field=-41nT, $\gamma_{n}=2\pi\, 11.8\mathrm{MHz T^{-1}}$, $\gamma_{e}=2\pi\, 28\mathrm{GHz T^{-1}}$, $\lambda=500$, $q=\frac{4}{2-4/(3+|P^e|^2)}$ \Rb{} polarisation decay time = 3.3ms, \Xe{} polarisation decay time = 10\,s. We assume pumping of the alkali fully saturates the polarisation and that the spin exchanging collisions between species are balanced.}
    \label{fig:response}
\end{figure*}

Figure~\ref{fig:response} shows the polarisation over a measurement cycle (i.e. two pump + decay periods) once the stable state has been reached (i.e. \pn{} transients have decayed). The simulation parameters were chosen to be similar to the experimental realisation that will follow in section \ref{expt}. In particular, we are working away from the compensation point, where the bias field strength is less than the sum of the atomic magnetisations. In this regime the precession frequency of the alkali is significantly faster than that of the noble gas, so that the dynamics of \pe{} and \pn{} aren't resonantly coupled.

% Want to get the idea in here that \pn{} is spending most of its time at its extremum positions- it's not spending most of its time doing the moving back and forth.
The data in Fig.~\ref{fig:response}(a-d) show the evolution of both \pe{} and \pn{} projected onto the $x-y$ plane.  In each case, \pn{} alternates between two points related by a $\pi/2$  rotation about the $z$-axis (which points out of the page), as it reacts to the alternating magnetic pulses. During the measurement of the alkali precession, \pn{} remains nearly constant. The alkali atoms, on the other hand, begin each decay period with \pe{} aligned to the $z$-axis due to the optical pumping, and then undergo free precession.

% How does \pn{} behave?
Let us first examine the behaviour of \pn{}. Comparing the response to an applied magnetic field ($B_{x,y}$) and a rotation ($\Omega_{x,y}$), we see that \pn{} is virtually indistinguishable, up to a scaling factor. The difference in size is just down to the values of the magnetic fields and rotations used in this simulation, which were chosen to give a similar polarisation response for the alkali. In fact, all the arcs formed by \pn{} in the this figure have the same shape, and vary only in scale and angle. This is because the arcs subtend a fixed angle, as determined by the magnetic pulses.
During the measurement time while \pe{} undergoes precession, \pn{} changes only a small amount, enough to offset the forwards and backwards arcs, but not enough to break the apparent symmetry between the four cases. 

% How does \pe{} behave?
The response of \pe{}, however, is clearly distinguishable in the four cases shown. For  a rotation, the driving of the alkali atoms is dominated by the noble gas magnetisation and bias field. In the case of an applied magnetic field, however, the alkali atom's direct response to the magnetic field is also significant due to their large gyromagnetic ratio. The result is a clear differentiation between magnetic and non-magnetic signals.

% The ramifications of the different properties of the species
The difference between properties of the noble gas and alkali atoms is crucial for this system to work. The magnetisation of the alkali atoms is small owing to the lower number density, it decays rapidly due to the short spin lifetime and it evolves quickly under the influence of a magnetic field to the large gyromagnetic ratio. The noble gas, on the other hand, has strong magnetisation due to the high number density, has a long-lived spin lifetime and responds more slowly to magnetic fields due to gyromagnetic ratio that is $\sim$1000 times smaller.  Consequently, the fast evolution of the alkali atoms is strongly influenced by the orientation of the noble gas.  The noble gas, however, does not respond strongly to the fast evolution of the alkali atoms.
There is a long term accumulation of \pn{} in the $z$-direction due to spin exchanging  collisions, but this is just a function of the average value of the alkali polarisation that is dominated by the short periods where it is pumped along the $z$-axis.

In an experiment we project only one axis of \pe{} into the optical polarisation of the probe beam. In this example we have measured the $x$-axis polarisation and plotted it in Fig.~\ref{fig:response}(e). Looking at only the first pulse response, up to 10\,ms, the traces look quite similar, but looking at the combined response over two alternate magnetic pulses we see responses for each of the rotations and magnetic fields that are substantially different. For the small signals normally measured with these devices, our system is linear, so that any combination of $x$ and $y$ rotations and magnetic fields can be found using linear regression over the four `signature' traces shown in Fig.~\ref{fig:response}(e). By performing this fit for each measurement cycle, our method can be used to extract time series for the four signals. This processing is not intensive and can be done in real-time with minimal computational resources.

\section{Experimental realisation}\label{expt}
We have conducted a proof of principle experiment to demonstrate the feasibility of our scheme. The cell used was approximately spherical and 1cm in diameter. Its nominal room-temperature contents where 0.01\,atm of \Xe{}, 0.4\,atm of N$_2$, and a drop of \Rb{}. This was heated to 125 degrees Celsius inside a four-layer mu-metal shield (Twinleaf MS1) with built in 3-axis magnetic coils.  The pump light was provided by a tapered amplifier seeded by a 795\,nm distributed Bragg reflector laser (Photodigm). The current to the tapered amplifier was pulsed to form the pump pulses. This provided around 2\,W peak power while on, with a leakage of the seed laser of approximately 10\,$\mu$W with the current off. The beam was enlarged to be similar to the cell size at roughly 10\,mm $\times$ 10\,mm.  The pump pulses were 300\,$\mu$s and applied every 2\,ms. After  several minutes of equilibrating time,  there was a $\sim$100\,nT effective field on the alkali due to \Xe{} polarisation. The probe light was provided by a 780\,nm external cavity diode laser, detuned from the D2 line by $\sim$100\,GHz and with a power of 2\,mW.  In these conditions, the T$_2$ times of \Rb{} and \Xe{} were measured to be near 2\,ms and 3\,s respectively.

The $z$-axis magnetic pulses were provided by a small, 5-turn, hand-wound pair of Helmholtz coils. These were 15\,mm diameter, with 7.5\,mm separation, and inside the oven. They were driven by a high-bandwidth audio amplifier (Yamaha A-S701) that was connected to the coils via two cascaded pairs of crossed diodes to prevent adding amplifier noise during precession. The amplifier could provide peak currents of up to 10\,A with pulse widths down to a few $\mu$s. A microcontroller was used to provide the pulse timing and apply small magnetic field modulations. To record the polarimeter signal, it was digitized at 14-bit and streamed at 125\,kS/s to a computer.

To set the operating point, a simple procedure was followed. The bias field was tuned to give approximately half of a Larmor period within the 1.7\,ms free decay window. This resulted in a bias field significantly below the compensation point. At this point, a signal was observed due primarily to residual transverse fields and beam misalignments. This was minimized by applying small field corrections and beam pointing corrections. 
The PANCo magnetic pulses were then turned on. To set the pulse amplitude, a 100\,pT transverse field was applied while the pulse amplitude was increased from zero until the difference in signal between alternate windows was similar in size to the signal in each window.

\begin{figure}[b!]
    \centering
    \includegraphics[width=0.7\linewidth]{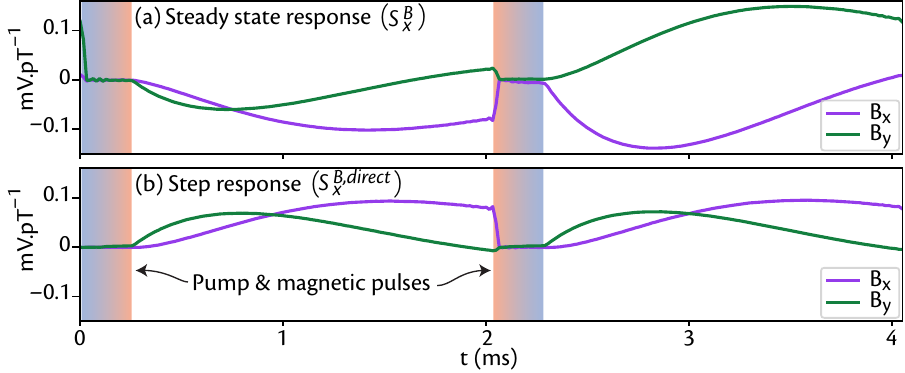}
    \caption{The polarisation of the \Rb{} ensemble measured in response to magnetic fields perpendicular to the pump axis. As in Fig.~\ref{fig:response}(e), this data is measured over one cycle of the alternating magnetic pulses.  a) Response to steady magnetic fields. These curves provide us with the magnetic field responses $S^{B}_{x,y}$, similar to those modelled in Fig.~\ref{fig:response}. b) Response immediately following application of a magnetic field step. As described in the text, these curves allow the extraction of a proxy for non-magnetic (rotational) signals in our data. These signatures were measured using $\approx 100\,pT$ amplitude square wave driving along the x- and y-axes of the system.}
    \label{fig:fast_slow}
\end{figure}

To extract measurements from polarimeter data, we need the signatures that represent the system's linear response to an external drive as shown in Fig.~\ref{fig:response}(e). We will denote the precession signals $S^{B}_{x,y}$ and $S^{\Omega}_{x,y}$ to be the signature curves for $B_{x,y}$ and $\Omega_{x,y}$ respectively.
These are non-trivial to obtain from theory as they depend strongly on experimental parameters, including imperfections, so instead we aim to measure them. 

The signatures for $B_{x,y}$ are obtained easily since the coils in the oven allow precise application of magnetic fields. The measured curves for $S^{B}_{x,y}$ are shown in Fig.~\ref{fig:fast_slow}(a). As we are lacking a precision rotation stage it is not possible to similarly extract $S^{\Omega}_{x,y}$. 
Instead, we use the response to a magnetic square wave to extract a proxy for rotational signals. 

We can consider the response of \pe{} to a magnetic field to be the sum of two components, $S^B_{x,y} = S^{B,direct}_{x,y} + S^{B,n}_{x,y}$, where the two terms on the right are the alkali's direct response to the applied field and that due to the field of the noble gas, respectively. We may first find $S^{B,direct}_{x,y}$ by measuring \pe{} immediately following a fast magnetic step, before the noble gas can respond.
Next, we recall that the system response to a rotation is dominated by the movement of \pn{}. Since \pe{} does not care what has caused displacement of \pn{}, there must be a relationship between $S^{B,n}_{x,y}$ and $S^{\Omega}_{x,y}$. In fact, in the regime we consider here, they are simply scaled versions of each other, as can be seen from Fig.~\ref{fig:response}(a,b) and the associated discussion. Thus, as a proxy for rotation or other non-magnetic signals we use the signatures given by $S^{B,n}_{x,y} = S^B_{x,y} - S^{B,direct}_{x,y}$. These proxy signatures are not calibrated, although we note that a more complete analysis of a magnetic step change has been used to extract a calibrated rotational response of a CW CSS \cite{PhysRevResearch.6.013339}. 

Based on the above considerations, we can now make use of the data from square wave modulation to extract our rotational proxies. At the end of the square wave, once the magnetic field has been stable for sufficient time, the experiment will be measuring $S^B_{x,y}$, the steady state magnetic signatures, which are shown in Fig.~\ref{fig:fast_slow}(a). Immediately after the step change in magnetic field, on the other hand, the alkali will respond only to the applied magnetic field, because the nuclear spin had not yet precessed due to its slow response. The precession of \pe{} after the step change therefore gives us $S^{B,direct}_{x,y}$, which are shown in Fig.~\ref{fig:fast_slow}(b).   We can then combine these data to find  $S^{B,n}_{x,y}=S^B_{x,y} - S^{B,direct}_{x,y}$, which as noted above, are proportional to the rotational signatures $S^{\Omega}_{x,y}$. We will refer to signals extracted using this signatures as $\beta_{(x,y)}$.

\begin{figure}[t!]
    \centering
    \includegraphics[width=0.7\linewidth]{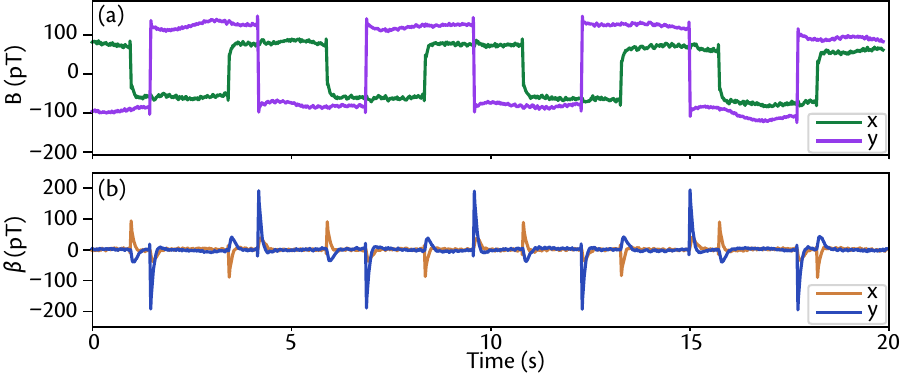}
     \caption{
    Magnetic and non-magnetic signals retrieved while square-wave magnetic modulation is applied on two axes. The modulation amplitude was 70~pT along $x$ and 100~pT along $y$. Part (a) shows the accurately extracted magnetic signal found by fitting $S^B_{x,y}$, although we note that there is significant background drift. Part (b) shows the $\beta$ signal extracted using $S^{B,n}_{x,y}$ for the non-magnetic response. This has spikes when the magnetic fields change rapidly, but is suppressed quickly afterwards.}
    \label{fig:B_mod}
\end{figure}

\begin{figure}[b!]
    \centering
    \includegraphics[width=0.7\linewidth]{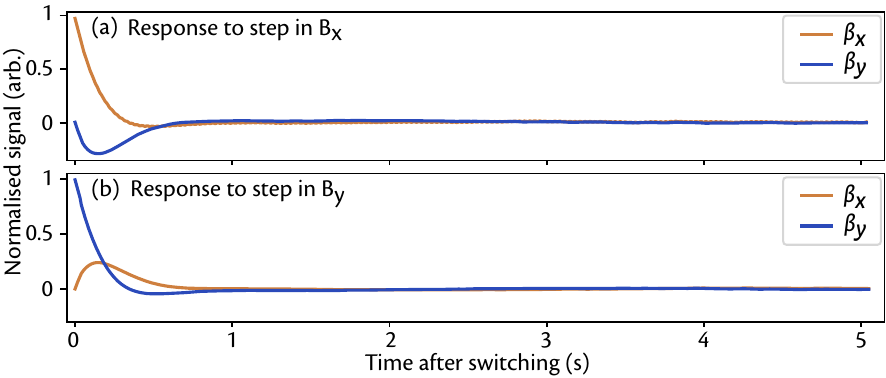}
     \caption{Data showing the suppression of magnetic response in the rotational proxy $\beta$.  Here we average over 70 step changes in applied magnetic fields in the $x$-direction (a) and $y$-direction (b).}
    \label{fig:suppression}
\end{figure}

The results of using our measured and inferred signatures are shown in Fig.~\ref{fig:B_mod}. Square wave magnetic modulations were applied on each axis. Every 4\,ms, pairs of precession periods were recorded and then fitted using the signatures. In Fig.~\ref{fig:B_mod}(a) we show the fitted amplitudes of $B_x$ and $B_y$, giving the expected square wave modulation. In Fig.~\ref{fig:B_mod}(b) we show the fitted amplitudes $\beta$ that correspond to rotational or non-magnetic signals. These curves show spikes in response to the rapid change in magnetic field before the noble gas has time to respond. This is qualitatively the same behaviour seen in CW comagnetometers. 

By averaging over many step-change modulations we obtain Fig.~\ref{fig:suppression}. This data shows that $\beta$ is insensitive to steady state magnetic fields. Compared to the initial response that follows the magnetic step, after 4 seconds the signal is suppressed by at least a factor of 180 on both axes, near the limit of the available signal-to-noise. The key conclusion from this data is that our method achieves strong suppression of magnetic field noise in the rotational response.

\begin{figure}[t!]
    \centering
    \includegraphics[width=0.7\linewidth]{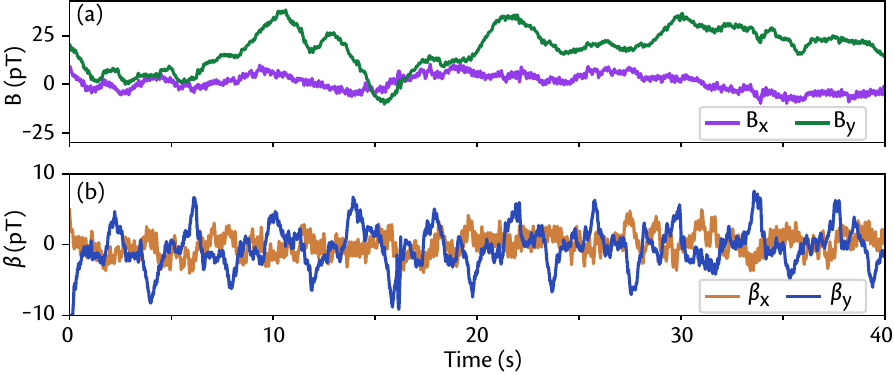}
     \caption{Measured signals while the optical table is wobbled to cause rotations at a level of $\sim 100\,\mu$Hz. The extracted magnetic signals in (a) show a large magnetic background due to a noisy current source. This noise is suppressed in the non-magnetic signals shown in (b).  The wobble of the optical table in the $y$-axis is seen clearly in the non-magnetic signal confirming that rotations are captured by our method with low levels of magnetic crosstalk.}
    \label{fig:wobble}
\end{figure}

To demonstrate that $\beta$ does actually capture rotations, we applied a periodic modulation of our optical table by driving a wobbling motion with a stepper motor pushing against the laboratory wall. From the geometry we estimate a peak rotation rate on the order of 100$\mu$Hz in the $y$-axis. This was carried out shortly after the measurements in Fig.~\ref{fig:B_mod} and used the same signatures for fitting. The results are shown in Fig.~\ref{fig:wobble}. The magnetic response in (a) shows drift in the B-fields of the order of 25\,pT.
The $\beta$ signal shows distinct oscillations in the $y$-axis that correspond to the periodic driving  of our optical table. We note that the magnetic background drift is not present in $\beta$, nor is the modulation in $\beta$ present in $B$. This indicates that any cross-talk is below the signal-to-noise ratio of our experiment.

While we emphasise that this demonstration is proof of principle only, we have shown that the system works as modelled in terms of distinguishing fields and enables moderate rotational sensitivity in a noisy background. Using this data we estimate a value for magnetic--rotation cross-talk of $0.2 \pm 0.1\mu$Hz$/$pT. 
For CW \SCC{}s, it is common to quote a suppression factor for magnetic response relative to an ideal SERF magnetometer. For comparison, we calculate that our measured cross-talk is equivalent to a suppression factor of 160 for an ideal self-compensated K-$^3$He comagnetometer,  similar to the best published for that configuration. This should be improved in future experiments with better characterisation of the rotational response. Similarly, the bias field, gas mixture and precession measurement times were chosen for convenience, so there is much room for optimisation.

\section{Idealised Sensitivity}
We now consider the limits of the PANCo scheme independent of experimental issues, as an estimate of how well the scheme might perform when applied to state-of-the-art systems that typically use K-$^3$He. We also consider optimising performance in terms of bias field.

We will assume that sensitivity is limited by white detection noise, allowing us to compare a standard CW \SCC{} under the the same conditions. While simplified, this gives an indication of how the scheme performs when limited by detection noise (such as photon or spin shot noise) that affects both systems the same way. In particular, we might expect some degradation of sensitivity for the PANCo case because it aims to retrieve four signals instead of just one. 
 
For the cell, we consider a hypothetical K-$^3$He cell with parameters similar to published results \cite{Kornack2002,Kornack2005}. We let the peak magnetisation of the alkali be $\lambda M_a =9\,$nT corresponding to a cell at $\approx 200\degree{}$C. To make the simulations tractable, we also assume the limit of long \He{} spin lifetime. This means we may consider a constant pre-polarised \He{} magnetisation (with $\lambda M_n P_n=100\,$nT) and neglect spin exchange and destruction terms which will be negligible over the time scale considered. Explicitly we set $R^{en}_{se}= R^{ne}_{se} = R^n_{sd} = 0$.

\begin{figure}[b!]
    \centering
    \includegraphics[width=0.7\linewidth]{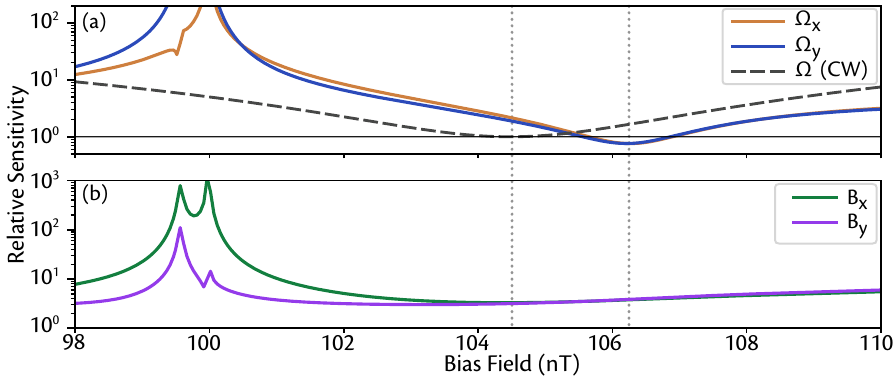}
     \caption{PANCo Sensitivity to rotations and magnetic fields. (a) Comparison of the PANCo rotational sensitivity as a function of bias field under white noise conditions. The sensitivity is compared to a single axis CW \SCC{} with the same operating parameters. The sensitivities have been normalised to the best performance of the CW \SCC{}, which occurs at the self-compensation point where the bias field is 104.5\,nT, as marked by a dotted line.
    The PANCo achieves a sensitivity that is 30\% better than the CW \SCC{} at bias of 106.3\,nT, also marked with a dotted line. We note that the model does not consider magnetic noise, which is only suppressed by the CW \SCC{} at the self-compensation point. (b) The PANCo scheme measures transverse magnetic fields simultaneously with rotation.  The magnetic sensitivity has been normalised to the performance of a SERF alkali magnetometer that was modelled by removing the noble gas and setting the bias field to zero. At 106.3\,nT, the magnetic sensitivity of PANCo is a factor of four worse than the ideal alkali magnetometer. The peaks near 100\,nT are due to the alkali seeing near-zero field such that the 4 signatures are no longer distinguishable.}
    \label{fig:sensitivity}
\end{figure}
 
To realise the CW simulation, we set the pumping rate to the alkali spin-destruction rate  $R_p = R^e_{sd} = 50$\,Hz, to yield a steady alkali polarisation of $P^e_z = 0.5$. For the PANCo simulation, we set the pump interval at $\tau = 20\,$ms and the pumping time to be $1$\,ms, giving a free precession time of 19\,ms. The pumping rate is enough to achieve $ P^e_z=0.99$ polarisation in that 1\,ms, while the mean alkali polarisation across a measurement period is $\langle P^e_z\rangle = 0.66$. The magnetic \PC{} pulses are applied simultaneously with the optical pumping and give a pulse area for the \He{} nuclear spin of $\pm\pi/2$. For both schemes we simulate a range of bias fields, although the CW scheme only suppresses magnetic noise at the compensation point.

For each bias field, we repeat the simulation twice: once with a small $x$-direction magnetic field, $B_x = 10^{-4}$\,nT, and once with a small rotation  $\Omega_x=10^{-4}\,$Hz. The simulations are run for 40\,s, such that the nuclear spin transients have had time to die out. Then, the final pair of decay traces, that is \pe{}  for the last $2\tau$ period, is taken as the `signature' for an applied field. The simulation does not have to be repeated in the orthogonal axis, as these results can be obtained by symmetry:  the $y$ component of \pe{} when being driven by a field in the $x$-direction is the same as the $x$ component when being driven by a field in the $y$-direction. 
With the four signatures from the two simulations, we then calculate the sensitivity achievable when fitting these parameters under the assumptions of ordinary least squares. Specifically, we find a representative covariance matrix for the four signals in each simulation as the inverse of a Fisher information matrix, which is in-turn constructed from the overlap integral between the four signature curves. The relative sensitivities between simulations are taken as the square-root of the diagonals of the covariance matrix. To provide reference sensitivity levels for each signal, we also model two single-axis CW situations: For rotations we model a CW \SCC{} under the same conditions and range of bias fields. For magnetic fields, we remove the noble gas from the model and turn off the bias field to simulate an optimal SERF magnetometer. To ensure consistency with the pulsed case, the CW cases were analysed using an identical procedure. However, for these there is only a single `signature' which is a DC offset corresponding to the CW signal size.

The results for measurements of rotations (and other non-magnetic signals) are summarised in Fig.\ref{fig:sensitivity}(a). The data shows that the PANCo scheme has very similar maximum sensitivity to a CW \SCC{}, despite measuring four parameters at once. In fact, at optimum bias field, the sensitivity is about 30\% better than the CW \SCC{}. Secondly, the sensitivity is optimised at a slightly higher bias field, $106.3\,$nT compared to  $104.5\,$nT for the CW \SCC{} case. 
The optimal bias field in both cases corresponds to where the maximum \He{} response occurs. The difference in optimum bias field is consistent with the higher average polarisation $\langle P^e_z \rangle$ in the pulsed PANCo scheme. Note that the optimal sensitivity values don't depend strongly on the polarisation. In particular, increasing the pumping rate for the CW case
%to 95\,s$^{-1}$
to match the mean \pe{} polarisation in the PANCo case reduces the optimal sensitivity slightly ($\sim$10\%). This is presumably because the decreased alkali coherence is partly balanced by an enhanced \pn{} response.

Our scheme also provides dual axis magnetic field measurements. The idealised sensitivity of the magnetic read-out is shown in Fig.~\ref{fig:sensitivity}(b). The performance in this case is normalised to an alkali SERF magnetometer. This shows that at the bias that provides the best rotational performance, PANCo sacrifices a factor of about four in magnetic sensitivity. The magnetic response is also seen to be relatively insensitive to changes in bias field for values near the compensation point.

\section{Discussion}

Our scheme has some potential advantages over other methods with respect to gyroscopy and exotic physics searches. Rotational sensing is a key part of inertial navigation systems, but it may be augmented with magnetic navigation systems \cite{Li_Wang_2014}. A sensor that combines both capabilities on two axes may prove useful. With respect to searches for new physics, our method has the potential to enable stronger rejection of magnetic noise, which is a key limiting factor for CW \SCC{}s.

Working towards these applications will require careful consideration of the magnetic cross-talk. The PANCo scheme can theoretically achieve zero cross-talk between magnetic and rotation signals, although this will only hold true when the signatures are known perfectly. Achieving low cross-talk in practice means good measurement of the signatures in the first place, and then low drift of experimental parameters so the signatures do not change. A common issue in comagnetometers is drift of the compensation point due either to stray $B_z$ fields or drifting polarisation.
For the simulation at 106.3\,nT we calculate that the bias field must remain within $\pm0.2\,$nT to keep the cross-talk below $0.2\,\mu$Hz/pT. This value of cross-talk corresponds to a `suppression' factor of around 150 for an equivalent CW comagnetometer. 
This corresponds to a relative bias stability of 0.2\% which is stringent, but achievable. Further, drifts in $B_z$ are easily detected in our scheme via the alkali precession frequency (by introducing a small transverse field if necessary), thus enabling a method for stabilising the bias field \cite{PhysRevApplied.12.024017}. We also expect drifts in other parameters, such as the magnetization of the spins, although the use of pulsed pumping will help in distinguishing this and other deleterious effects.
It is notable that our experiment, albeit with a different mixture of gases, has already achieved a high level of magnetic suppression with only minimal optimisation. This gives us reason to be optimistic about the levels of suppression that can be achieved in future work.

The large magnetic pulses introduced in the PANCo scheme are also a significant potential noise source.  We expect that variations in pulse strength will cause variations in the signatures, and thus cross-talk between magnetic and rotation signals similar to variation in bias field. In addition, these large pulses could induce eddy currents in nearby conductors such as magnetic shields, which may in turn cause magnetic noise that persists during the free precession period, and indeed the induced response in shields can approach 1\,ms. This could vary due to changes in pulse size or external effects such as temperature changes, and so cause real and unwanted variations in magnetic fields.  At present neither of these effects seem dominant in our experiment: we conclude this from a simple test with 10$\times$ weaker pulses (area of $\pi/20$, obtained by turning down the amplifier). This gave no corresponding reduction in low frequency magnetic noise floor, and only a degradation in rotational noise floor due to degraded magnetic suppression. We note that the coils used at present are deliberately small compared to the size of the shield (7.5\,mm radius compared to $\approx 10$\,cm distance to shields), resulting in a relatively small $\sim 100$\,nT field at the shields during the pulses. We expect that future works will need careful design of coils to balance homogeneity on the sample while reducing eddy currents in the shields, in addition to careful tuning and control of the magnetic pulses. It is perhaps salient to note that the application of phase-controlling pulses are used in quantum information processing and NMR for dynamic decoupling \cite{Souza_2012,Wolfowicz2016,reif2021solid}. While the purpose of the pulses in our system is different, there is a wealth of experimental and theoretical work that shows pulse sequences can be used in quantum limited systems with great success.

A final noise source we expect in future experiments is angle or `pointing error' caused by fluctuations in the directions of the optical fields. It has been shown that operation at the compensation point can aid pointing suppression in pulsed systems \cite{Wang2024phd}. 
Preliminary modelling of PANCo suggests that pointing errors will give rise to their own unique signatures in the precession signals. If we are able to characterise and fit these signatures to our data, our method may enable elimination of pointing error from magnetic and rotational data away from the self-compensated regime.  We expect that the extraction of signatures for pointing error will also be a subject for future work.

Before probing the sensitivity limits outlined above, modifications will be required to our proof-of-principle experiment to address sources of noise and drift. Immediate improvements will come from replacing the current source used to drive the transverse coils.  Following this, improvements to the stability of our optical beam paths to reduce pointing noise, pump beam inhomogeneity, optimising the gas cell mixtures and addressing the impact of field gradients will form a part of future experimental work.

\section{Conclusion}
Through the applications of magnetic and optical pulses, we have devised a new protocol for extracting signals from an atomic comagnetometer. The scheme is able to measure magnetic and non-magnetic effects on two orthogonal axes with a noise that, in ideal conditions, should be comparable to the state-of-the-art continuous wave scheme, where only a single parameter is measured. A proof-of-principle experimental realisation based on \Rb{} and \Xe{} was presented with promising results. In particular, we demonstrated that the PANCo scheme effectively suppresses cross-talk between magnetic and non-magnetic signals to a level that is comparable to the best self-compensated alkali-noble comagnetometers. This result was achieved with minimal optimisation, yet the scheme has a wide parameter space for performance gains in terms of the applied pulses, the applied bias fields, gas mixtures and investigations of techniques to diagnose and minimise other technical noise. In future work we plan to explore these factors to uncover the practical limits of the scheme.\\

\noindent\textbf{Acknowledgments}\\
This project was funded by the Australian Research Council (DP240100534) and ANU Connect Ventures (DTF405). We thank John Close and Derek Kimball for helpful discussions.\\

\end{document}